\documentclass[twocolumn, showpacs, superscriptaddress, nofootinbib,aps, prl, 10pt, floatfix]{revtex4-1}
\usepackage[utf8]{inputenc}
\usepackage[english]{babel}
\usepackage{graphicx}
\usepackage{amsmath}
\usepackage{mathtools}			%for underbrackets
\usepackage{bm}
\usepackage{environ}
\usepackage{comment}
\usepackage{grffile}
\usepackage{xcolor}
\usepackage{booktabs}			% for nice tables
\usepackage{placeins}			% floatbarrier
\usepackage{amsfonts}
\usepackage{amssymb}
\usepackage{wasysym}
\usepackage{dsfont}
\usepackage{color,soul}
\usepackage{ulem}				% Underlining and strikeouts
\usepackage{dcolumn}			% Align table columns on decimal point
\usepackage{bm}				% bold math
\usepackage[pdftex, colorlinks, citecolor=blue, linkcolor=blue, urlcolor=blue]{hyperref}
\usepackage[all]{hypcap}			% makes hyperref work OK
\usepackage{paralist}			% for the \compactitem command
\usepackage[version=4]{mhchem}	% for chemical reactions
\usepackage{tikz-cd}				% for the square of chemical reactions
\usepackage{chemfig}

\definecolor{myred}{rgb}{0.8,0.1,0.1}
\definecolor{mygre}{rgb}{0.0,0.5,0.2}
\definecolor{myblu}{rgb}{0.0,0.0,0.9}
\graphicspath{{./figures/}}

\begin{document}

\title{Slow fluctuations shape cell signaling near a bifurcation in embryonic stem cells}
\author{Fiorella Fabris} 
\affiliation{Instituto de Investigaci\'on en Biomedicina de Buenos Aires (IBioBA) – CONICET/Partner Institute of the Max Planck Society, Polo Cient\'{i}fico Tecnol\'ogico, Godoy Cruz 2390, Buenos Aires C1425FQD, Argentina}
\affiliation{Research and Technology Institute, AREA Science Park, Trieste 34149, Italy}
\author{Dhruv Raina}
\affiliation{Department of Systemic Cell Biology, Max Planck Institute for Molecular Physiology, Otto-Hahn-Str.~11, D-44227, Dortmund, Germany}
\author{Christian Schröter}
\affiliation{Department of Systemic Cell Biology, Max Planck Institute for Molecular Physiology, Otto-Hahn-Str.~11, D-44227, Dortmund, Germany}
\affiliation{Leiden Academic Centre for Drug Research, Leiden University, Einsteinweg 55, 2333CC Leiden, The Netherlands}
\author{Luis G. Morelli}		
\affiliation{Instituto de Investigaci\'on en Biomedicina de Buenos Aires (IBioBA) – CONICET/Partner Institute of the Max Planck Society, Polo Cient\'{i}fico Tecnol\'ogico, Godoy Cruz 2390, Buenos Aires C1425FQD, Argentina}
%
%\affiliation[2]{}
%\affiliation[3]{}
%
\date{\today}
\begin{abstract}
%% Abstract length limits: about 5% of article length and < 500 words for all journals except PRL, which allows ≤ 600 characters.
Excitable and oscillatory responses are widespread in cell signaling, but how these arise in the same system is unclear. We propose a minimal model of a saddle node bifurcation on an invariant circle, where slow fluctuations modulate transitions between excitable and oscillatory states, and noise drives excitability. Fitting the model to cell signaling data from embryonic stem cells reveals that these cells operate close to the bifurcation. Proximity to the bifurcation allows the cell to tune different dynamic states, suggesting a general principle for versatile cell signaling.
\end{abstract}
\maketitle

%%%%%%%%%%%%%%%%%%%%%%%%%%%%%%%%%%%%%%%%%%%%%%%%%%%%%%%%%%%%%%%
% Our focus: the signaling modes are still an open question. 
% Some say stochastic, maybe excitable, some say oscillatory. 
% Here we show: maybe both. 
%%%%%%%%%%%%%%%%%%%%%%%%%%%%%%%%%%%%%%%%%%%%%%%%%%%%%%%%%%%%%%%

\vspace{0.50cm}
\noindent
% Cells process the signals they sense from the environment through signaling networks, generating dynamic activity patterns.
%
Cells process environmental signals through signaling networks that generate dynamic activity patterns.
These activity patterns relay the information to downstream transcriptional programs that guide cellular responses such as cell survival, fate decisions, and collective behaviors in tissues.

Prominent examples of dynamic signaling are pulses and oscillations observed across diverse biological contexts, 
even under constant external stimulation~\cite{casani2022, levine2013, purvis2013}. 
%
% Pulsatile and oscillatory signaling dynamics have been increasingly observed across diverse biological contexts, even under constant external stimulation~\cite{casani2022, levine2013, purvis2013}. 
% and are emerging as a central layer of temporal organization in the cell~\cite{casani2022, levine2013, purvis2013}. 
%NOTCH
% Notch signaling has been reported to generate excitable dynamics in mouse cultured cells from the segmentation clock~\cite{hubaud2017}, while persistent oscillations were observed in zebrafish cells~\cite{webb2016, rohde2024}.
%
In cultured cells from the mouse segmentation clock, 
Notch signaling was reported to generate excitable dynamics~\cite{hubaud2017},
while persistent oscillations were observed in zebrafish cells~\cite{webb2016, rohde2024}.
%
% NFKB
% in zebrafish the stimulus is not constant (waves)
% Similarly, the transcription factor NF$\kappa$B displays distinct dynamic activity patterns depending on stimulus and context. 
%
% Under constant stimulation with TNF-$\alpha$, NF$\kappa$B exhibits continuous oscillations in immune cells, while bacterial LPS triggers a single prolonged wave in the same system~\cite{nelson2004, lee2009}. 
%
Similarly, the transcription factor NF$\kappa$B displays distinct dynamic activity patterns in immune cells, 
such as continuous oscillations or a single prolonged pulse depending on stimulus and context~\cite{nelson2004, lee2009}. 
In zebrafish hematopoiesis, NF$\kappa$B exhibits slower, temporally separated pulses~\cite{campbell2024}.
%
%
% p53
% maybe we can read Monke2017 and say something about isolated pulses as well?
% The pulse trains length depend on the stimulus, maybe say something also about that=
%
% Another example is the response of the tumor suppressor p53 to DNA damage, showing a range of dynamic modes depending on the cell line~\cite{stewart-ornstein2017}. 
%
Another example is the response of the tumor suppressor p53, showing a range of dynamic modes depending on the cell line~\cite{stewart-ornstein2017} and quantitatively encoding for the level of DNA damage in signaling parameters~\cite{gutu2023}. 
Within the same cell line, p53 displays signatures of excitability, featuring isolated pulses under basal conditions, and sustained oscillations at high stimulation levels~\cite{monke2017}.
Dynamic patterns determine cell fate, with oscillations favoring cell cycle arrest and DNA repair, while sustained signal activation promotes apoptosis or senescence~\cite{bonsignore2025,purvis2012}.
\begin{figure}[b!]
\label{fig:data}
\centering
\includegraphics[width=\columnwidth]{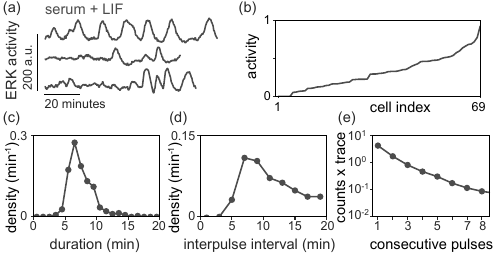}
\caption{ 
{\bf Intermittent oscillations of ERK signaling in embryonic stem cells.} 
(a) Representative traces of an ERK activity sensor, 
(b) activity is the fraction of time that single cells spent pulsing in the experiment, 
(c) pulse duration and  
(d) inter pulse intervals distributions, and 
(e) pulse sequence lengths counts.
Data replotted from~\cite{raina2022}.
} 
\end{figure}

% --------- Dyncode
% Another interesting example is the ERK signaling system. 
% \red{Say what ERK is and does.}
%
% Here we focus on ERK signaling~\cite{}. 
%
Here we focus on ERK activity dynamics, a key component of a signaling system controlling cell growth, differentiation, and death~\cite{lavoie2020, ram2023}. 
ERK dynamics has been studied in various cell systems upon short-term and continuous stimulation~\cite{ram2023}. 
% with EGF
%
Diverse dynamic behaviors have been reported, ranging from sporadic activity bursts~\cite{albeck2013} to stochastic pulsing~\cite{aoki2013}, and oscillatory behavior~\cite{nakayama2008, shankaran2009}.
%
%Under short-term and continuous EGF stimulation, ERK signaling has been studied in various systems and display diverse behaviors~\cite{ram2023}. 
%
%In many cell types, ERK activity is pulsatile and stochastic~\cite{aoki2013}. 
%
% Other cells show oscillatory ERK activity~\cite{nakayama2008, shankaran2009}, while some display sporadic pulses~\cite{albeck2013}.

Both in mouse early embryonic development and in mouse embryonic stem cells, FGF/ERK signaling controls cell fate decisions~\cite{schroter2015, ram2023b}. 
We recently reported that, in mouse embryonic stem cells, sustained exposure to FGF4 can induce intermittent oscillations of ERK activity~\cite{raina2022}, where oscillatory intervals alternate with silent periods and isolated pulses, Fig.~\ref{fig:data}(a). 
These intermittent oscillations were inferred from key features of the pulsing statistics, Fig.~\ref{fig:data}(b-e). 
We observed that single cell ERK activity, defined as the time the cell spent pulsing, is heterogeneous across the population, Fig.~\ref{fig:data}(b).
The time elapsed between the beginning and the end of a single pulse, termed pulse duration, 
reveals a well-defined timescale of approximately 7 min, Fig.~\ref{fig:data}(c). 
Considering the interpulse interval (IPI), the time between peaks of two successive pulses, we observed a long-tail revealing the presence of silent periods between pulses or pulse sequences, together with a modal value close to 7 min indicating consecutive pulses, Fig.~\ref{fig:data}(d). 
Consecutive pulses can occur in sequences, constituting a key feature of ERK signaling dynamics, Fig.~\ref{fig:data}(e).
Taken together, these features provide a characterization of ERK signaling dynamics in embryonic stem cells. 
%
%Taken together, these features encompass ERK signaling dynamics signatures in embryonic stem cells.
%
Since signaling dynamics may determine cellular responses, here we aim to identify the mechanism that shapes them. 
Simple stochastic pulsing models failed to capture the pulsing statistics observed in the experiments~\cite{raina2022}. 
To address this gap, in this work we introduce a generic dynamical framework capable of capturing these intermittent oscillatory dynamics.

\begin{figure}[t]
\label{fig:model}
\centering
\includegraphics[width=\columnwidth]{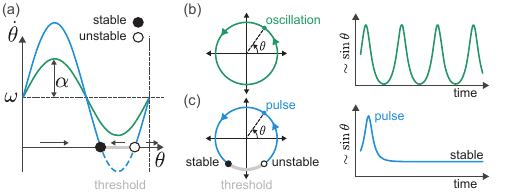}
\caption{
{\bf A general theory of a SNIC bifurcation allows for both oscillatory and excitable dynamics.}
% {\bf A general theory allowing for both oscillatory and excitable dynamics.}
(a) Phase portrait shows how instantaneous phase velocity is modulated by a sinusoidal. Depending on the relative values of $\alpha$ and $\omega$, the model can generate either 
(b) sustained oscillations, or
(c) isolated pulses.
(a)-(c) Cases $\alpha<\omega$ (green) and $\alpha>\omega$ (blue) are illustrated.
}
\end{figure}

\vspace{0.50cm}
\noindent
{\textit{Theory.---}}
%
% As a general framework, we consider the Adler equation for phase locking~\cite{adler46}, $\dot\theta = \omega + \alpha \sin (\theta)$, a generic model for the Saddle-Node on invariant circle (SNIC) bifurcation~\cite{strogatz}, Fig.~\ref{fig:model}. 
%
As a general framework, we consider the Adler equation for phase locking~\cite{adler46}, $\dot\theta = \omega + \alpha \sin (\theta)$, a generic model for the Saddle-Node on invariant circle (SNIC) bifurcation~\cite{strogatz}, Fig.~\ref{fig:model}.
%
% \begin{equation}
% \dot\theta = \omega + \alpha \sin (\theta) + \sqrt{2D} \xi(t)
% \end{equation}
%
Here $\theta$ is a phase variable taking values in the unit circle. 
The levels of active ERK could be mapped from this phase by a $2\pi$-periodic non-linear mapping, for example the sine function of the phase. 
The instantaneous phase speed $\dot\theta$ has contributions from an autonomous frequency $\omega$, and a phase dependent  modulation of amplitude $\alpha$, Fig.~\ref{fig:model}(a).
Depending on parameters, the model can produce both sustained oscillations and single pulses, Fig.~\ref{fig:model}(b, c). 
The Adler equation has been applied to Josephson photonic devices~\cite{danner2021}, superconducting circuits~\cite{markovic2019}, quantum dots~\cite{liu2016}, lasers~\cite{paulau2012, cerjan2014}, optomechanical oscillators~\cite{amitai2017, yamaguchi2026}, ratchets and overdamped motions in harmonic washboard potentials under constant forcing~\cite{reimann2002, hanggi2009}, and as a canonical model in neuroscience~\cite{roulet2016}, among others.
%
%
% Within this framework, we will consider two different hypotheses. 
%
% Fig.~\ref{fig:model}(b).
%
% For illustration, here we will use the sine of the phase as a non-linear mapping to obtain some oscillatory concentration, for example ERK, levels, from the phase variable.
%
%

Within this framework, we will consider two different hypotheses. 
%
% Pulses and pulse trains could result from excitable dynamics.
%
Pulses and pulse trains could result from excitable dynamics.
For $\alpha>\omega$, there is a pair of stable and unstable fixed points in the unit circle, Fig.~\ref{fig:model}(a, c). 
The separation between these two states defines a threshold $\Delta = \left| \pi - 2 \arcsin{(\omega/\alpha)} \right|$.
% Fig.~\ref{fig:model}(a, c).
%
% Pulses and pulse trains could result from excitable dynamics, generated by an additive white noise term 
%
Then, excitability can be generated by an additive white noise term 
\begin{equation} \label{eq:excitable}
\dot\theta = \omega + \alpha \sin (\theta) + \sqrt{2D} \xi(t)
\end{equation}
with phase diffusion coefficient $D$ and uncorrelated fluctuations, $\langle \xi(t) \xi(t') \rangle=\delta(t-t')$.
%
% For $\alpha>\omega$, there is a pair of stable and unstable fixed points in the unit circle, Fig.~\ref{fig:model}(a, c). 
%
%FF
In this excitable model, noise perturbs the system around the stable fixed point. If noise is strong enough, it can kick it beyond the unstable fixed point and trigger an excursion around the circle, representing a single pulse, Fig.~\ref{fig:model}(c).
%Bef: Noise perturbs the state about the stable fixed point, and if the strength of the noise is large enough, it can kick the state beyond the unstable fixed point and trigger an excursion around the circle, representing a single pulse, Fig.~\ref{fig:model}(c).
%
For larger noise strength, pulses are produced at higher frequencies.
While the resulting pulses occur irregularly in time, coherence resonance~\cite{pikovsky97, lindner2004} may enhance pulsing regularity, generating noisy oscillatory dynamics.
% at which the pulsing regularity is enhanced, and may resemble noisy oscillatory dynamics.
%
% Excitable model with noise, or slow bifurcation dynamics of the threshold.
%
% It may be that the subthreshold model, with noise, is able to describe the noisy pulsing, and that stochastic resonance allows for more regular pulsing providing the observed statistics of pulse trains.

As an alternative, pulses and pulse trains could result from transitions between silent and oscillatory states, caused by a slow fluctuation of the threshold.
This could be described by a stationary Ornstein-Uhlenbeck (OU) process governing the dynamics of the amplitude $\alpha(t)$~\cite{gardiner}, 
\begin{align} 
\dot\theta(t) &= \omega + \alpha(t) \sin (\theta(t))   \label{eq:threshold}  \\
\dot\alpha(t) &= - \frac{1}{\tau} \left( \alpha(t) - \alpha_0 \right) + \frac{\sqrt{2\sigma}}{\tau}  \eta(t)  \label{eq:ou}
\end{align}
where 
$\tau$ is the timescale of the OU process, 
$\alpha_0$ is the mean amplitude,
$\sigma$ sets the strength of the instantaneous amplitude fluctuations, 
and we do not include a white noise term in phase dynamics.
%
% Here, phase dynamics does not include a white noise term, and it is the slow fluctuation of $\alpha(t)$ that produces excursions on one side or the other of the bifurcation separating the oscillatory state from the steady state. 
%
In this fluctuating threshold model, the system will oscillate while $\alpha(t)>\omega$, Fig.~\ref{fig:model}(a, b), and single pulses would be produced as short excursions of $\alpha(t)$ into the oscillatory state.

\vspace{0.50cm}
\noindent
{\textit{Results.---}}
To evaluate these two hypotheses, we fit the models to data from~\cite{raina2022} by means of an approximate Bayesian computation (ABC)~\cite{marjoram2003, sisson2007, toni2009, csillery2010}. 
The ABC approach performs a search in parameter space while optimizing a cost function that measures the distance between simulated and experimental traces in the chosen statistical features, see End Matter.
This distance captures the statistics of pulse duration, IPI, pulse rate and consecutive pulses.
With these observables, we aim to capture the timescales of pulse and silence durations, global activity and pulse sequences.
%
% This distance captures the statistics of pulse duration and IPI distributions, since together they encompass information about the duration of pulses and silences.
%
% We also require the fit to approximate the pulse rate distribution, which informs about global pulsing activity in a trace, and the distribution of pulse train lengths, to capture the longer timescale of pulse trains.
%
% We run the ABC algorithm for at least XYZ iterations, with minimum tolerance $\epsilon = XYZ$. 
%
% The tolerance is the median of the distances between a set of simulated time series and the experimental data. 
%
%
% Supp. Figs. of epsilons and fit results.

\begin{figure}[t]
\label{fig:wn.ou}
\centering
\includegraphics[width=\columnwidth]{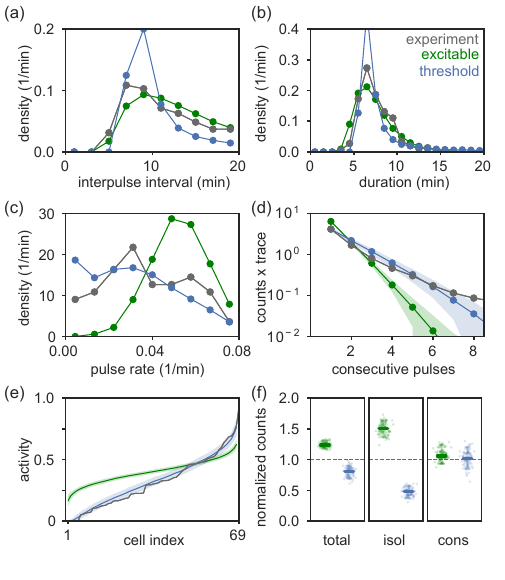}
\caption{{\bf White noise and threshold fluctuations individually capture complementary aspects of experimental data.} 
Experiments (grey) and simulations of the excitable (green) and fluctuating threshold (blue) models.
Statistics of 
(a) interpulse intervals,
(b) pulse durations, 
(c) pulse rates, and 
(d) consecutive pulse sequences,
(e) single cell activity and 
(f) the count of total, isolated and consecutive pulses.
Counts in (f) are normalized to experimental values (dashed grey line).
} 
\end{figure}

The excitable model Eq.~(\ref{eq:excitable}) provides a good fit to the IPI and pulse duration distributions, Fig.~\ref{fig:wn.ou}(a, b).
However, it fails to capture other aspects of the data, such as the pulse rate distribution and the length of the pulse sequences, Fig.~\ref{fig:wn.ou}(c, d).
Similar to other simple stochastic models~\cite{raina2022}, it underestimates the frequency of long consecutive pulse sequences, and does not produce traces with low pulse rates, favoring higher pulse rates.
%
%In contrast, the fluctuating threshold model Eq.~(\ref{eq:threshold}) improves the fit of the pulse rate and length of the pulse sequences distributions, Fig.~\ref{fig:wn.ou}(c, d), but makes a slightly poorer description of IPI and pulse duration distributions, Fig.~\ref{fig:wn.ou}(a, b).

% The fluctuating threshold model Eq.~(\ref{eq:threshold}) makes a slightly poorer description of IPI and pulse duration distributions, Fig.~\ref{fig:wn.ou}(a, b). 
%
In contrast to the excitable model, the fluctuating threshold model Eq.~(\ref{eq:threshold}) 
underestimates the long tail in the IPI distribution 
and has a sharper pulse duration distribution, Fig.~\ref{fig:wn.ou}(a, b). 
However, it substantially improves the fit of the pulse rate and length of the pulse sequences distributions, Fig.~\ref{fig:wn.ou}(c, d).
%
%
%To assess the scope of the fit, we compared observables not included in the fit cost function: the single cell activity, and the total, isolated and consecutive pulse counts. 
%
% Furthermore, the excitable model fails to describe the variability observed in single cell activity, and the fluctuating threshold model provides a very accurate description of this feature, which was not included in the fit cost function, Fig.~\ref{fig:wn.ou}(e).
%
Furthermore, the fluctuating threshold model provides a very accurate description of single cell activity, which was not included in the fit cost function, Fig.~\ref{fig:wn.ou}(e).
%
% Both models generate consecutive pulse pairs in agreement with experimental data, but fail to match the count of total and isolated pulses, Fig.~\ref{fig:wn.ou}(f).
%
Altogether, the fluctuating threshold model provides a better description of the data. 
Still, the timescale required to produce long pulse sequences is incompatible with the shorter excursions required to produce single pulses, resulting in underestimation of isolated pulses, Fig.~\ref{fig:wn.ou}(f).
%
% Both models generate consecutive pulse pairs in agreement with experimental data, but fail to match the count of total and isolated pulses, Fig.~\ref{fig:wn.ou}(f).
%
%
%
\begin{figure}[t]
\label{fig:dual}
\centering
\includegraphics[width=\columnwidth]{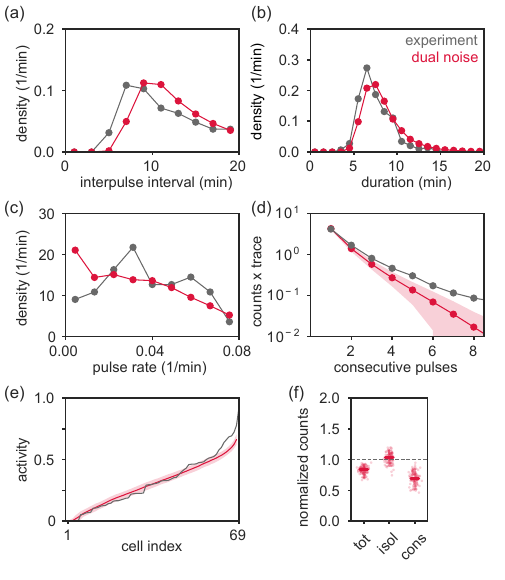}
\caption{
{\bf Dual noise theory captures intermittent oscillations statistics from experiments.}
Fits and results for model combining white noise model and slow fluctuating threshold (red) together with experimental data for comparison (grey), showing
(a) interpulse intervals,
(b) pulse durations, 
(c) pulse rates, and 
(d) consecutive pulse sequences,
(e) single cell activity and 
(f) count of total, isolated and consecutive pulses.
Counts in (f) are normalized to experimental values (dashed grey line).
} 
\end{figure}

% \vspace{0.50cm}
% \noindent
%{\bf Threshold fluctuations with additive noise.}
%
% Partial failure of individual components, Fig.~\ref{fig:wn.ou}.
%
% Taken together, the results from the fits of both models in Fig.~\ref{fig:wn.ou} indicate that each hypothesis is better at capturing complementary aspects of the data, but neither of them can individually capture the full complexity of ERK dynamics in embryonic stem cells. 
%
% Taken together, these results indicate that each hypothesis is better at capturing complementary aspects of the data, but neither of them can account for the full complexity of ERK dynamics in embryonic stem cells. 
%
Taken together, these results indicate that each hypothesis captures complementary aspects of the data, suggesting that the two components together might be able to improve the description of intermittent ERK oscillations.
%
% HERE write the interpretation of results, reasons why one is better than the other for each observable.
%
%FF commented the following paragraph
%\red{The excitable model provides a better description of IPI statistics and pulse duration variability, and the fluctuating threshold model is better at capturing long pulse sequences, pulse rates, and the variability in single cell activity.}
%
% These results suggest that the two components together might be able to capture the intermittent ERK oscillations.
%
Thus, we propose a dual noise model that combines additive white noise phase fluctuations causing excitability with slow fluctuating threshold dynamics,
%
%
%
% \FloatBarrier 
\begin{equation}
\dot\theta(t) = \omega + \alpha(t) \sin (\theta(t)) + \sqrt{2D} \xi(t) \, ,
\end{equation}
where the amplitude dynamics is given by Eq.~(\ref{eq:ou}).
We used the ABC approach to fit this model to the data, see End Matter.
This dual noise model combines features of the excitable model capturing the IPI statistics and pulse duration variability, Fig.~\ref{fig:dual}(a, b), together with the fluctuating threshold capacity to generate traces with low pulse rates and traces with long pulse sequences, Fig.~\ref{fig:dual}(c, d).
%
% Moreover, this dual model captures the single cell activity plot and the statistics of total, isolated, and consecutive pulses, Fig.~\ref{fig:dual}(e-g).
%
Moreover, the dual model captures observables that were not included in the fit, Fig.~\ref{fig:dual}(e-f).
%
% Moreover, the dual model provides a good description of quantities that were not included in the fit, like the single cell activity variability and the count of total and isolated pulses, and consecutive pulse pairs, Fig.~\ref{fig:dual}(e, f).
%
%
%Thus, we conclude that both ingredients present in the dual model are necessary to capture the full complexity of the data. 
%
%the slow threshold fluctuations are necessary to produce longer pulse sequences and the high variability in single cell activity, and the additive phase noise is necessary to reproduce the long tail of the IPI distribution and the variability in pulse duration.
%
% These two components together also compensate the opposing discrepancies in the total pulse count and the isolated pulses count observed in the models for the individual components.

\begin{figure}[t!]
\centering
\label{fig:bifurcation}
\includegraphics[width=\columnwidth]{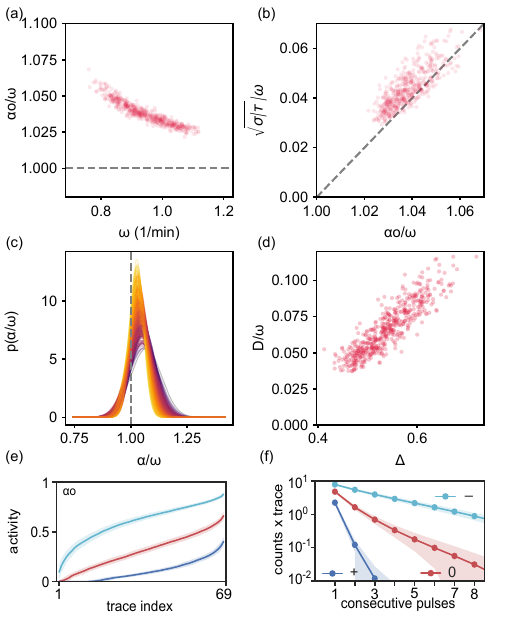}
\caption{
{\bf Parameter values position the system close to the SNIC bifurcation.}
(a) Values of $\alpha_0/\omega$ vs. $\omega$ for all parameter sets accepted in the fit for the dual noise model. Dashed grey line indicates $\alpha_0/\omega = 1$.
(b) Standard deviation vs. mean of the OU process Eq.~(\ref{eq:ou}), scaled by $\omega$. Dashed grey line is $\alpha_0/\omega-1$.
(c) Probability densities for the OU process Eq.~(\ref{eq:ou}) for all parameter sets from the fit. 
The color bar indicates the means $\alpha_0/\omega$, ranging from $1.021$ (yellow) to $1.068$ (purple).
% $1.0216176045683687$ to $1.068396486786191$.
(d) Scaled phase diffusion coefficient vs. threshold for all parameter sets from the fit. 
(e) Activity and 
(f) consecutive pulse sequences from 100 realizations of numerical simulations of the dual noise model with parameter values from the best fit (red), and changing the value of $\alpha_0$ by a 10\% fold increase (blue) and reduction (light blue).
} 
\end{figure}

% \vspace{0.50cm}
% \noindent
%{\bf Parameter values are close to the bifurcation.}
%
% The ABC fitting algorithm returns probability densities for the values of the model parameters.
%
The ABC fitting algorithm returns an ensemble of parameter sets that minimize the cost function.
The plot of $\alpha_0/\omega$ vs. $\omega$ reveals that all results are consistently above $\alpha_0/\omega = 1$, Fig.~\ref{fig:bifurcation}(a).
This means that the system sits, on average, in the excitable regime in all the accepted parameter sets for the dual noise model.
Furthermore, while dispersion along the $\omega$ axis is large, 
the distance to $\alpha_0/\omega=1$ line is relatively small and has low variability.
Whereas the actual values of $\alpha_0$ and $\omega$ seem to be less constrained, its ratio appears to be critical to recover the statistics in the data. 
Thus, the fit indicates that the system operates in the vicinity of a critical point 
where a bifurcation between excitable and oscillatory dynamics occurs.
%
% Thus, the fit indicates that the system appears to sit close to the bifurcation, and while the actual values of $\alpha_0$ and $\omega$ seem to be less well determined, its ratio appears to be critical to recover the statistics in the data. 
%
% \blu{Solution to the OU process.}
%
The steady state of the OU process of Eq.~(\ref{eq:ou}) is a Gaussian distribution with mean $\langle \alpha \rangle = \alpha_0$ and variance 
$\langle (\alpha - \alpha_0)^2 \rangle = \sigma/\tau$~\cite{gardiner}.
% $\sigma_{\alpha}^2 = \langle (\alpha - \alpha_0)^2 \rangle = \sigma/\tau$.
%
Taking the parameter sets from the fit, we can compute the standard deviation of the steady state distribution of $\alpha$ values. 
Scaling by the frequency $\omega$ to render these quantities comparable, we find that the standard deviation 
falls above the distance to bifurcation $\alpha_0/\omega-1$ and is positively correlated with the mean, Fig.~\ref{fig:bifurcation}(b). 
That is, the steady state distributions $p(\alpha/\omega)$ become broader as the mean increases and moves away from the bifurcation point $\alpha_0/\omega=1$, Fig.~\ref{fig:bifurcation}(c). 
Taking the integral of these distributions up to $\alpha/\omega = 1$ we obtain the mean probability for the system to be in the oscillatory state is $0.19$ with standard deviation $0.03$.
This means that the system is in the oscillatory state about 20\% of the time, and given the relatively small variability, this value is consistent across parameter sets obtained from the fit.
%
% This means that the system is most of the time in an excitable state, spending about 20\% of the time in the oscillatory state, and this value is very consistent across parameter sets obtained from the fit.
%
% np.std(cdf)
% Out[48]: 0.029   ~ 0.03
% np.mean(cdf)
% Out[49]: 0.193   ~ 0.20 0.19 
%
Similarly, the scaled phase diffusion coefficient $D/\omega$ grows with increasing threshold $\Delta$ separating the stable and unstable fixed points, Fig.~\ref{fig:bifurcation}(d).
This means that noise strength is larger in parameter sets that are further away from the bifurcation, allowing for similar excitability rates.
Additionally, the variance of the OU process and the phase diffusion coefficient have similar values, and are distinctly different from zero, Fig.~\ref{fig:bifurcation}(b, d). 
This indicates that the fit assigns similar weights to the two noise components. 
Thus, we conclude that both ingredients present in the dual model are necessary to capture the full complexity of the data.

Due to this proximity to the bifurcation, the statistics of activity and pulse sequence length are sensitive to variations in $\alpha_0/\omega$, Fig.~\ref{fig:bifurcation}(e, f).
%
%Due to this proximity to the bifurcation, variations in $\alpha_0/\omega$ can control the statistics of activity and pulse sequence length, Fig.~\ref{fig:bifurcation}(d, e), while variations in other model parameters display a much weaker sensitivity.
%
We previously reported that the prevalence of intermittent oscillations increases with FGF concentration in FGF mutant cells~\cite{raina2022}.
These results suggest that FGF levels could control the proximity to the bifurcation, eventually crossing it.
%
%
% \red{LM: This results suggest that FGF levels could control the proximity to the bifurcation, eventually crossing it.... in the discussion relate this to the role of Notch in mouse segmentation clock cells.}
%
%
%
% \vspace{0.50cm}
% \noindent
% {\bf Parameter sensitivity.}
% %
% Previous experiments revealed that ERK signaling dynamics is affected by the levels of extracellular FGF.
% %
% By varying FGF concentration in FGF mutant cells that could not produce the ligand.
% %
% It was reported that the prevalence of intermittent oscillations increases with FGF concetration.
% %
% We wondered which aspects of the theory could account for such changes.
% %
% So we do a parameter sensitivity study to see how all parameters in the model affect key features as variability in activity plots and the pulse sequence statistics, Fig.~\ref{fig:sensitivity}.
%
% \vspace{0.50cm}
% \noindent
% {\bf Parameter sensitivity.}
%
%

\vspace{0.50cm}
\noindent 
{\textit{Discussion.---}}
% 
% Here we show... 
Here we introduced minimal description that captures essential features of previously reported intermittent ERK oscillations in ESCs~\cite{raina2022}. 
Using a framework based on a  generic model of the SNIC bifurcation~\cite{strogatz}, we explored two different mechanisms: additive white noise driving excitability, and a slowly fluctuating frequency modulation driving oscillations.
We show that, combined, these two mechanisms account for key experimental observables of intermittent oscillations.
Additionally, statistical features of the data are best described when the dynamical system operates close to a critical point that separates excitable and oscillatory regimes. 
This dynamics could potentially arise from multiple nested negative feedback loops present in ERK signaling system~\cite{lake2016, ram2023}.

In the model, slow fluctuations control the position of the system relative to the bifurcation and thereby enables the emergence of long pulse trains. 
The length of these pulse trains is controlled by the average phase amplitude modulation.
We previously showed that FGF levels control the length of pulse trains, suggesting that FGF may act by tuning the proximity to the bifurcation. 
A similar role was attributed to Notch signaling in the mouse segmentation clock~\cite{hubaud2017}.
Additionally, we reported that pulsing rate decays across the cell cycle~\cite{raina2022}. 
This could either indicate that the system moves further away from the critical point with cell cycle progression, or that fluctuations could decrease with increasing cell volume.

% Operating in the vicinity of the critical point allows the system to flexibly switch between signaling modes, providing sensitivity to external signals.
% %
% Other signaling systems operate near the transition between oscillatory and non-oscillatory behavior~\cite{camalet2000, eguiluz2000, westendorf2013, webb2016, hubaud2017}. 
% %
% In some cases, the non-oscillatory state has been interpreted as an excitable regime producing isolated pulses \cite{martinezcorral2018, monke2017, hubaud2017}, similar to what we observe in ERK signaling in embryonic stem cells. 
% %
% This suggests that positioning near the bifurcation may be a general principle to generate versatile dynamic patterns of activity across diverse biological contexts. 

A small number of cellular signaling systems regulate the bewildering number of possible cellular responses~\cite{li2019}.
The ERK signaling system, which is highly conserved across species, displays a rich repertoire of short-term dynamics: single‑cell studies have revealed that it can produce transient, pulsatile, or oscillatory responses in multiple cellular contexts~\cite{ram2023}. 
This diversity contributes to the regulation of a variety of cell-specific processes such as proliferation or differentiation~\cite{ram2023b}. 
While mathematical models of ERK signaling typically focus on single dynamical behaviors~\cite{ram2023} , here we considered a minimal description that provides a generic framework for intermittent oscillations. 
Our framework provides a foundation for developing a unified description of ERK dynamics across cellular contexts, highlighting how dynamic positioning near bifurcations may modulate multiple cellular responses.
More broadly, different cellular contexts may be represented by different positions in parameter space, thereby generating different types of signaling dynamics responses.
%
% In this way, context dependency may originate at the level of dynamical positioning through distinct dynamic signaling behaviors.

%%% NEW paragraph
It has been proposed that proximity to a bifurcation underlies enhanced sensitivity and signal amplification in diverse biological sensory systems, for example in hair cells of the inner ear~\cite{camalet2000, eguiluz2000, momi2025, alonso2025}, thermosensing in the snake pit organ~\cite{graf2024}, olfactory receptor neurons~\cite{choi2024}, actin dynamics in chemotactic sensing amoebae~\cite{westendorf2013}, and in receptor networks processing time-varying signals~\cite{stanoev2020}. 
In pattern formation, criticality has been linked to robust and adaptable tissue organization, as shown for rhizoid patterning in a liverwort~\cite{mercadal2026, garciaortiz2026}. 
In spatially extended systems, the excitable or oscillatory nature of local dynamics can shape the character of wave propagation~\cite{ditalia2022, franovic2026}. 
%
% giving rise to trigger waves or phase waves under different coupling regimes~\cite{ditalia2022, franovic2026}. 
%
SNIC bifurcations have been proposed in generic and geometric models of fate selection and patterning during embryonic development~\cite{jutrasdube2020, farjami2021}, and experimental observations during worm development have shown widespread gene expression oscillations arrest through a SNIC bifurcation upon hatching~\cite{meeuse2020}.
More generally, proximity to a SNIC bifurcation produces a ghost with a diverging characteristic timescale~\cite{strogatz}, and ghost channels and cycles have recently been proposed as building blocks of a general theory of biological computation~\cite{koch2024a, koch2024b, koch2025}. 
Together, these examples highlight that biological systems may tune their proximity to a bifurcation to switch between dynamical regimes, suggesting a general principle across biological scales.

\vspace{0.50cm}
\noindent
{\textit{Acknowledgements.---}}
%\section*{Acknowledgements}
% {\bf Acknowledgements}
% \red{We are grateful...}
%
We thank Roman Rolla and Gabriela Petrungaro for valuable comments and suggestions.  
This work was supported by ANPCyT grants PICT 2017 3753 and PICT 2019 0445 awarded to LGM and FOCEM-Mercosur (COF 03/11) to IBioBA. 
% ANPCyT PICT 2013 1301, PICT 2017 3753 to LGM,
%
FF was supported by a CONICET fellowship and LGM is a researcher of CONICET.
LGM acknowledges support from the ICTP through the Associates Programme (2026-2031), and thanks the QLS section at ICTP for warm hospitality.
FF acknowledges financial support under the National Recovery and Resilience Plan (NRRP), mission 4, component 2, investment 1.1, and call for tender no. 1409 published on 14 September 2022 by the Italian Ministry of University and Research (MUR), funded by the European Union — NextGenerationEU — CUP J53D23015070001. 
CS acknowledges funding from the Max-Planck-Gesellschaft.
%\newpage

%\bibliography{erk_theory.bib}
%\bibliography{dyncode,erk_theory}         
\bibliography{bibliography.bib}

% \vspace{0.50cm}
% \noindent

\mbox{}
\newpage
\cleardoublepage
\cleardoublepage
% \newpage
%
% Figures, tables, equations and pages in the supplement are numbered S1, S2 etc.
\renewcommand{\thefigure}{S\arabic{figure}}
\renewcommand{\thetable}{S\arabic{table}}
\renewcommand{\theequation}{S\arabic{equation}}
\renewcommand{\thepage}{S\arabic{page}}
\setcounter{figure}{0}
\setcounter{table}{0}
\setcounter{equation}{0}
\setcounter{page}{1} % not 0 as \newpage already started a supplementary page
% References continue the numbering from the main text.

\section{Supplementary Information}
\vspace{0.10cm}
% \noindent
\subsection{Fitting theories to experimental data}
\begin{figure}[t!]
\centering
\label{fig:epsilons}
\includegraphics[width=\columnwidth]{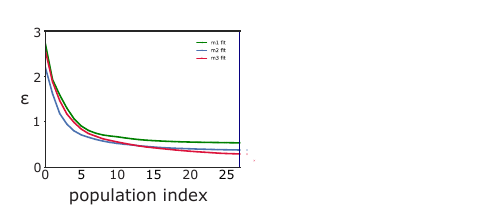}
\caption{
{\bf Convergence of the tolerance for the three fits performed.}
(a) Tolerance $\epsilon$ as a function of population index for the white noise (green line), threshold fluctuations (blue line) and dual noise models (red line).
} 
\end{figure}

We fitted models to the experimental data from~\cite{raina2022} using an Approximate Bayesian Computation (ABC) approach~\cite{marjoram2003,toni2009,csillery2010}, based on the Sequential Monte Carlo implementation~\cite{sisson2007, filippi2013} from \texttt{pyABC}~\cite{klinger2018}, with independent uniform priors (Table~\ref{tab:uniform_priors}) and multivariate Gaussian perturbation kernels~\cite{toni2009,toni2010}.
Each population consisted of $500$ accepted parameter sets. Convergence to the posterior distribution was monitored through a tolerance value, defined as the median distance between accepted simulations and the experimental data in the previous population. The simulations were stopped after $27$ population iterations~\cite{curacosta2021,toni2010}. 
The evolution of the tolerance $\varepsilon$ across populations shows convergence with this number of iterations, Fig.~\ref{fig:epsilons}.

We defined a distance $D$ based on summary statistics of intermittent oscillations, Fig.~\ref{fig:data}, including interpulse intervals (IPI), pulse durations (PD), pulse rates (PR), and pulse train statistics (PT):
\begin{equation}
D =
d_{\mathrm{IPI}}
+d_{\mathrm{PD}}
+d_{\mathrm{PR}}
+d^{*}_{\mathrm{PT}}.
\end{equation}
%
% Since fitting the full distributions is computationally demanding, 
Each contribution $d$ was defined as a normalized euclidean distance 
between quartiles $Q_1$, $Q_2$ and $Q_3$ of the simulated (S) and experimental (E) distributions, 
\begin{equation}
% d_{\mathrm{X}} =
d =
\frac{\sqrt{
(Q_{1\text{S}}-Q1_{1\text{E}})^2
+
(Q_{2\text{S}}-Q_{2\text{E}})^2
+
(Q_{3\text{S}}-Q_{3\text{E}})^2 }
}{Q_{3\text{E}}-Q_{1\text{E}}} .
\end{equation}
%
% with $\mathrm{X} = \{ \mathrm{IPI}, \mathrm{PD}, \mathrm{PR}, \mathrm{PT}\}$.
%
For the pulse train statistics we defined 
% term $d^{*}_{\mathrm{PT}}$,
%
\begin{equation}
d^{*}_{\mathrm{PT}}
=
\sqrt{
\sum_{i=1}^{8}
w_i
\frac{
\left[
\rho_\text{E}(i)-\rho_\text{S}(i)
\right]^2
}{
\rho_\text{E}(i)^2
}},
\end{equation}
where $\rho(i)$ denotes the number of pulse trains of length $i$. 
Weights $w_i$ are defined as
\begin{equation}
w_i=
\frac{1-i/8}{8-9/2}
\end{equation}
to assign more importance to shorter pulse trains, which are more abundant and therefore have more statistical weight. 
%
%\red{The distributions obtained for all fitted models are shown in Fig. XYZ.}
%
%
%
%
\begin{table}[ht!]
\centering
\begin{tabular}{|c|c|c|}
\hline
\textbf{parameter} & lower bound & upper bound \\
\hline
$\omega$ & $\pi/14\,\text{min}^{-1}$ & $8\pi/7\,\text{min}^{-1}$ \\
\hline
$\alpha_0/\omega$ & $0$ & $2$ \\
\hline
$\tau$ & $1$ & $200\,\text{min}$ \\
\hline
$\sigma$ & $0$ & $3\,\text{min}^{-3/2}$ \\
\hline
$D$ & $0$ & $2\,\text{min}^{-2}$ \\
\hline
\end{tabular}
\caption{Parameter ranges for uniform priors. 
The same ranges are used for all three models fitted.
In the white noise model Eq.~(\ref{eq:excitable}), 
the range of $\alpha/\omega$ is the same as $\alpha_0/\omega$.}
\label{tab:uniform_priors}
\end{table}

\vspace{0.10cm}
% \noindent
\subsection{Numerical simulations of phase trajectories}
Numerical phase trajectories were integrated using the Heun algorithm~\cite{sanmiguel2000} for the excitable model Eq.~(\ref{eq:excitable}), as well as for the OU slow amplitude fluctuation, Eq.(~\ref{eq:ou}). 
For phase trajectories involving OU amplitude dynamics Eq.~(\ref{eq:threshold}) and (\ref{fig:dual}) we used the modified Runge-Kutta scheme discussed in~\cite{sanmiguel2000}, which is computationally more expensive but converges correctly to Gaussian white noise in the small timescale limit. 
%
% For the dual noise model, we used the Runge-Kutta scheme~\cite{sanmiguel2000}, which is computationally more expensive but converges correctly to Gaussian white noise in the small timescale limit. 
%
Random numbers were generated using the Box-Muller-Wiener method~\cite{kloeden2003}. 
Time series were generated with integration time step $dt=0.001 \text{ min}$ over a total duration $T=1000 \text{ min}$. 
%
% Trajectories were sampled every $d=1$ steps. 
%
For the posterior analysis in Figs.~\ref{fig:wn.ou}, \ref{fig:dual}, and~\ref{fig:bifurcation}, we simulated $N=69$ time series with duration $T=120 \text{ min}$, consistent with the experimental data. 
Initial conditions for the phase variable were drawn from a uniform distribution $U(0,2\pi)$. 
For simulations including OU amplitude fluctuations, the initial amplitude was set to its mean value $\alpha_0$. 
%
% In models with OU amplitude fluctuations, fixed points were defined using the mean modulation amplitude $\alpha_0$.
%

\vspace{0.10cm}
% \noindent
\subsection{Pulse detection and summary statistics}
%
% Pulses were identified from the time series $X(t)=\sin\theta$. 
%
%First, local maxima and minima were detected. 
First, local maxima and minima were detected from the time series $X(t)=\sin\theta$. 
Candidate extrema were then required to exceed predefined amplitude thresholds $A_\text{th}=\pm 0.9$ for maxima and minima. 
In both cases, consecutive extrema of the same type were required to be separated by at least $W_\text{th}=100 \text{ frames}$. 
%
%To ensure consistent pulse boundaries, 
We imposed that only one minimum exists between consecutive maxima, and vice versa, retaining only the most prominent extrema within each interval. 
The maxima identify the pulses peak. 
We used the phase trajectory $\theta(t)$ to determine the beginning and the end of the pulse associated to each peak.
In the oscillatory regime, the boundaries were identified as the last crossing before the peak and the first crossing after the peak at phase $3\pi/2$.
In the excitable regime, we defined the beginning of the pulse as the last crossing of the unstable fixed point, that we identified by following the phase trajectory from the maximum to the preceding minima. 
Similarly, the end of the pulse was the first crossing of the stable fixed point, identified by following the phase trajectory from the maximum to the subsequent minima. 
In phase models with an OU process, fixed points in the excitable regime were computed using the mean amplitude $\alpha_0$. 
%
%In phase models with an OU process, in the excitable regime fixed points were computed using the mean amplitude $\alpha_0$, whereas in the oscillatory regime the boundaries were identified as the last crossing before the peak and the first crossing after the peak at phase $3\pi/2$.
%
% In the oscillatory regime, the boundaries were identified as the last crossing before the peak  and the first crossing after the peak at phase $3\pi/2$.
%
% whereas in the oscillatory regime the boundaries were defined at phases $\pm \pi$.
% phases $-\pi$ and $+\pi$.

Pulse duration was computed as the elapsed time between the pulse beginning and end. 
The IPI, as the time required to complete a phase advance of at least $2\pi$. 
A pair of pulses was consecutive if pulses were separated by an interval shorter than the mean duration of the two corresponding pulses, following~\cite{raina2022}. Then, a pulse was classified as consecutive if it belonged to a consecutive pulse pair, and isolated otherwise. 

\end{document}